\documentclass{article}

\usepackage{amsmath}
\usepackage{amssymb}
\usepackage{graphicx}

\begin{document}

\title{Dynamics of bacterial aggregates in microflows}
\author{Ana Carpio, Universidad Complutense de Madrid, Spain \\
Baldvin Einarsson, Air-Worldwide, Boston, USA, \\
David R. Espeso, Universidad Carlos III de Madrid, Spain}

\date{Dec 24, 2016}

\maketitle

{\bf Abstract.} Biofilms are bacterial aggregates that grow on moist surfaces. Thin homogeneous biofilms naturally  formed on the walls of conducts may serve as biosensors, providing information on the status of microsystems (MEMS) without disrupting them.  However, uncontrolled biofilm growth may largely disturb the environment they develop in, increasing the drag and clogging the tubes. To ensure controlled biofilm expansion we need to understand the effect of external variables on their structure. We formulate a hybrid model for the computational study of biofilms growing in laminar microflows. Biomass evolves according to stochastic rules for adhesion, erosion and motion, informed by numerical approximations of the flow fields at each stage. The model is tested studying the formation of streamers in three dimensional corner flows, gaining some insight on the effect of external variables on their structure.

\section{Introduction}
\label{p161:intro}

As the size of the components of technological devices diminishes, new procedures to measure their inner variables without disturbing the system must be developed. For some microdevices, cheap and environmentally friendly monitoring might be achieved exploiting the bacteria that live in them. Bioremediation policies already benefit from microorganisms. Bacteria feeding on a wide variety of toxic pollutants are deliberately released to clean up oil spills or to purify underground water in farming land and mines 
\cite{slimy}. For technological purposes, the ability of bacteria to emit optic signals is more appealing. Microorganisms naturally occurring in the environment fluoresce in response to the presence of certain chemicals or certain processes. Such is the case of bioluminiscence phenomena in the southern seas. 

Many bacterial species survive in moist environments forming aggregates called biofilms. Microorganisms adhere to surfaces, forming colonies and changing their phenotype to produce  extracellular polymeric matrix (EPS). This matrix shelters them from antibiotics, disinfectants, flows and external aggressions. Biofilms may be considered biological materials,
whose properties are governed by environmental factors affecting cellular behavior.
Recent attempts to engineer devices out of biofilms successfully produced electrooptical devices \cite{delorenzo1}. The advancement of synthetic biology is paving the way for the 
use of biofilms as bioindicators  or biosensors in the environment \cite{morin}. 
There are efforts to use biofilms emitting optic signals as microsensors in microdevices.  
Bacteria can be genetically engineered to change their color in response to variations in the environment. Properly modified, bacteria growing in the devices could give local information of the temperature or other variables, without perturbing the internal flow, since the typical size of bacteria is of the order of microns. To indicate the magnitude of variables on the surfaces they attach to,  biofilms should be homogeneous and thin. Pattern formation
may largely disrupt the environment they grow in. To be able to exploit bacteria in a controlled way, we must understand the influence of external factors on their collective dynamics. 
 

Biofilms are a mixture of living cells embedded in an exopolyscacharid
matrix which contains different kinds of metabolic by-products, that can be
generically considered as 'biomass'. In fact, the formation of biofilms in flows 
may be included in a more general group of physical processes where adhesion 
mechanisms drive agglomeration of matter to create different geometries.
The mechanical behavior of the biomass (EPS, cells, debris) and its interaction
with the flow seem to be relevant, allowing for growth of structures that do not align 
with the streamlines of the flow, but may cross the mainstream or wrap around 
tubes forming helices instead \cite{stonestreamer,helices}.  

In this paper, we propose a computational framework to study the growth of biological aggregates in flows triggered by adhesion of particles, much faster than growth due to nutrient consumption.
The biofilm is considered a  biomaterial with known average cohesive properties formed by a soft sticky matrix of EPS, debris, and other substances secreted by the cells included in it or floating around. We formulate  stochastic rules for biomass adhesion, erosion and motion informed by the continuous flow fields around the expanding aggregate, that are approximated by a finite difference discretization strategy using a fixed mesh  to reduce the computational cost.   The resulting model is tested studying biofilm streamer formation in laminar corner flows.

The paper is organized as follows. In section 2, we describe the general framework
and  collect the rules for biomass behavior.  Section 3 illustrates the numerical 
results and discusses the insight gained on the dynamics of the aggregates. 

\section{Hybrid description of biofilms in microflows}
\label{p161:hybrid}

Hybrid models combine continuous descriptions of some relevant fields, such as 
concentrations, flow fields or EPS matrix production, with discrete descriptions of the cells \cite{hybridmatrix,pre,hybridhydro}. The situation we examine here fits better as interaction of the surrounding fluid with a elastic biofilm structure whose growth is mediated by adhesion processes.  From a computational point of view,  biomass is considered as a mixture of bacteria and organic matter allocated on a grid which may behave in different ways in response to external conditions  with a certain probability.  

\begin{figure}
\centering
\includegraphics[width=5cm]{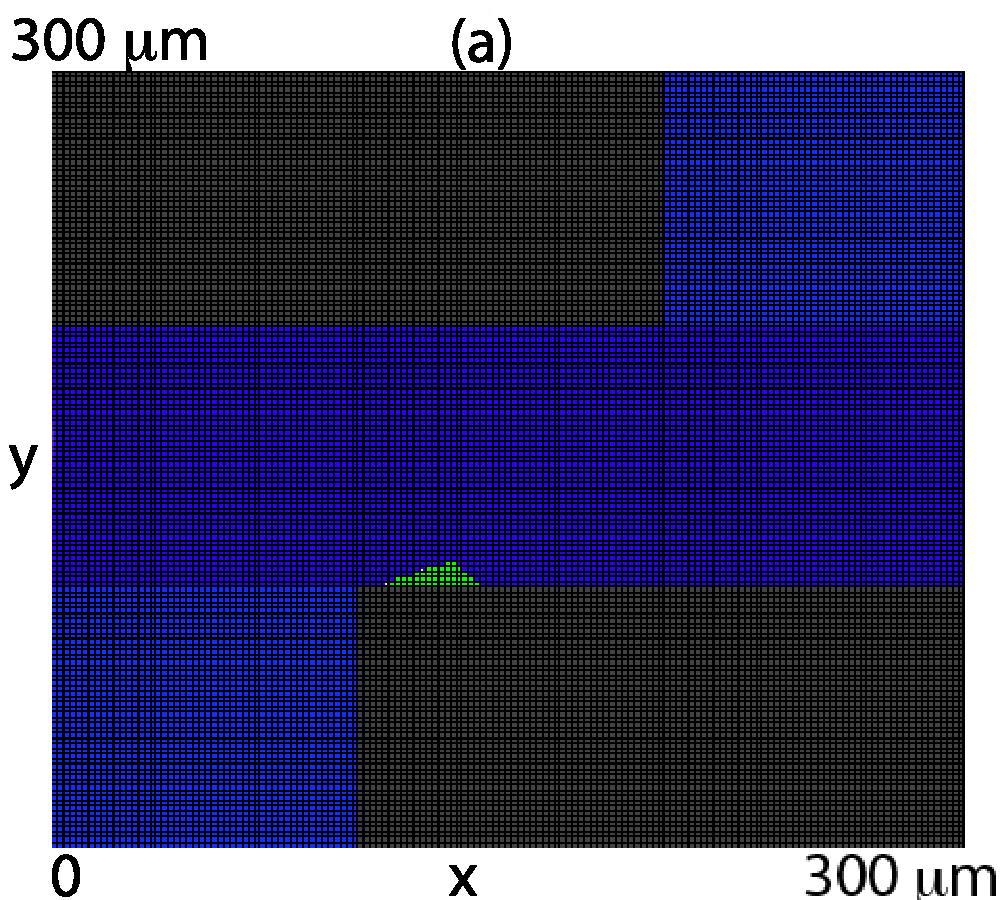}
\includegraphics[width=5.5cm]{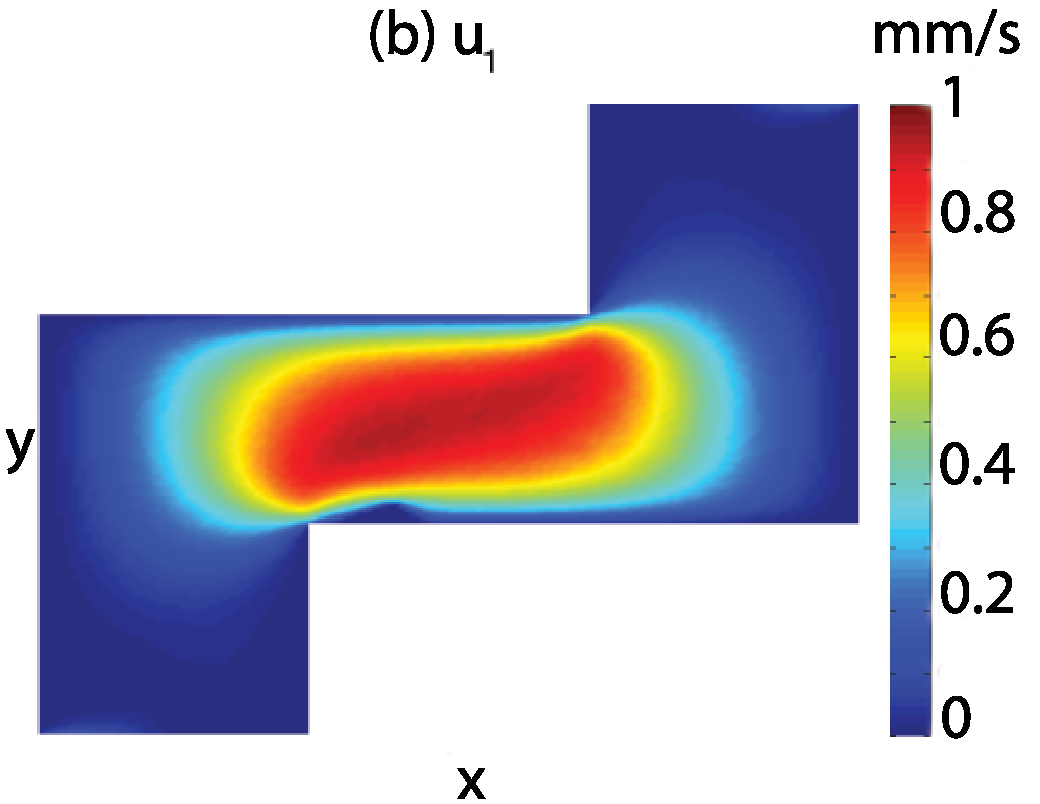}\\
\includegraphics[width=5.5cm]{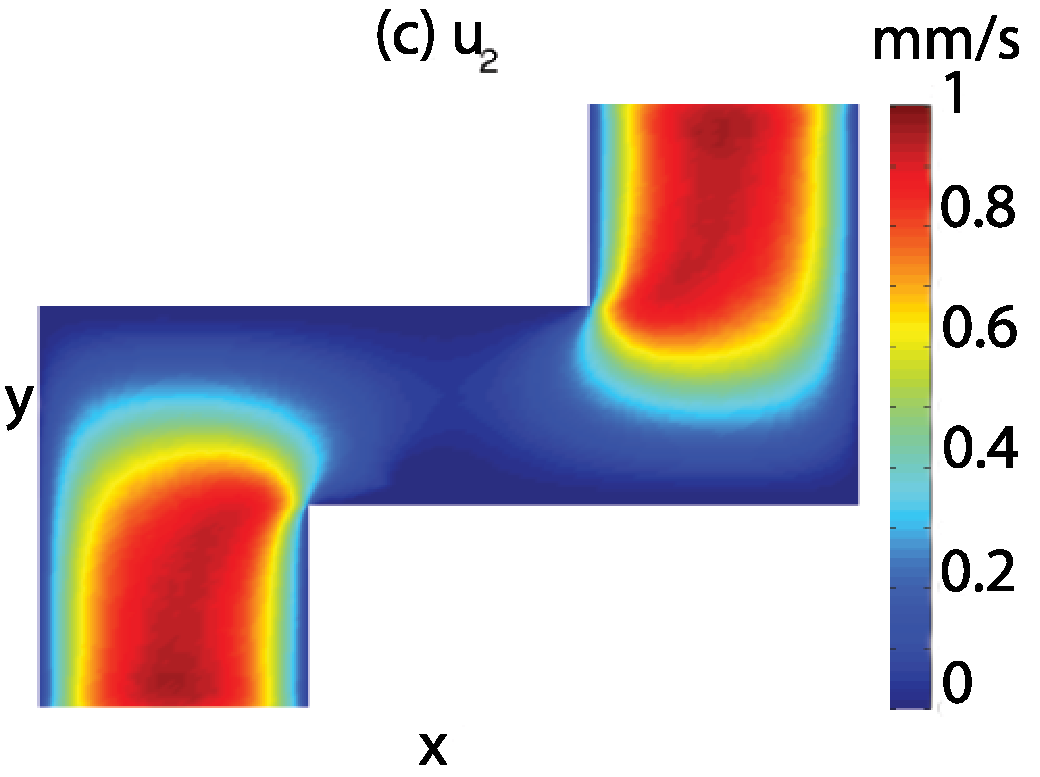}
\includegraphics[width=5.5cm]{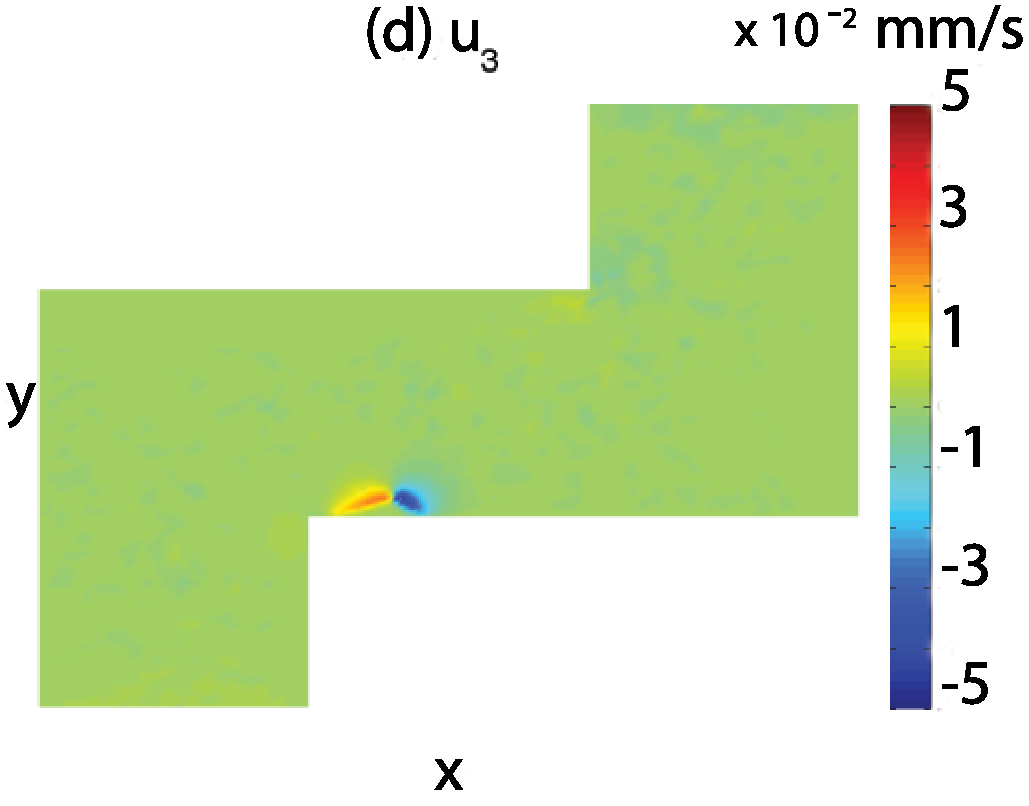}\\
\includegraphics[width=5.5cm]{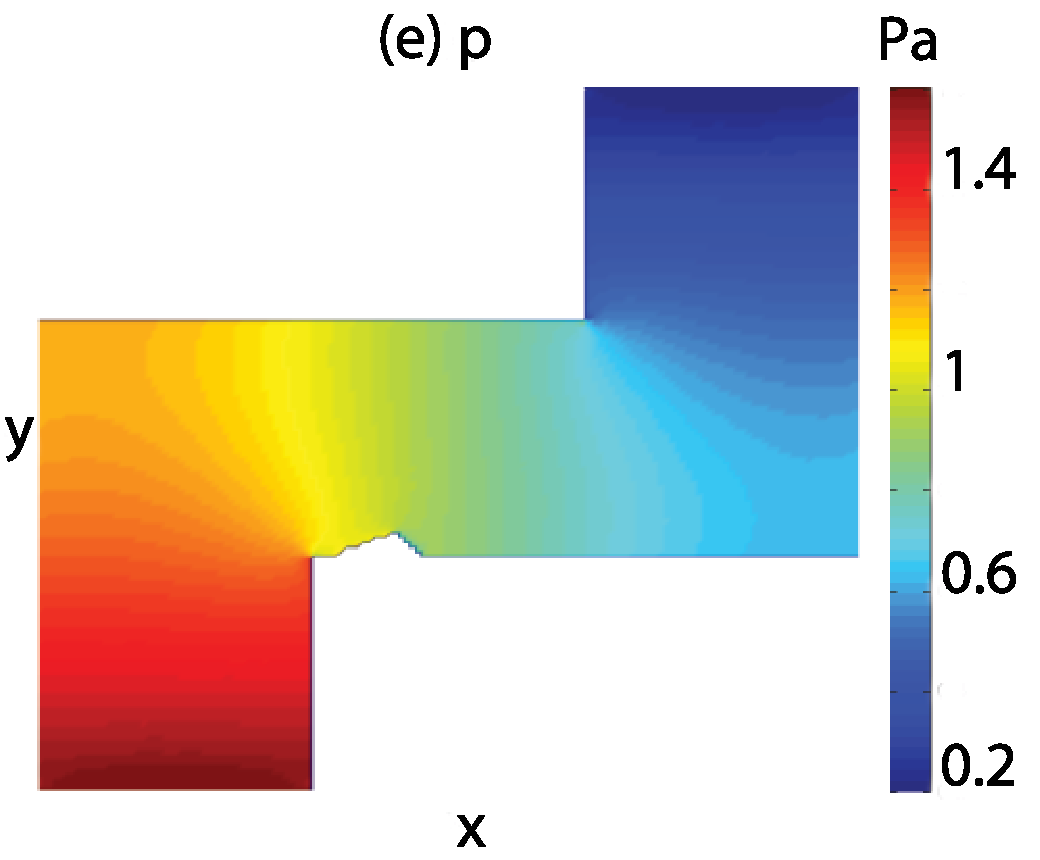}
\includegraphics[width=5.5cm]{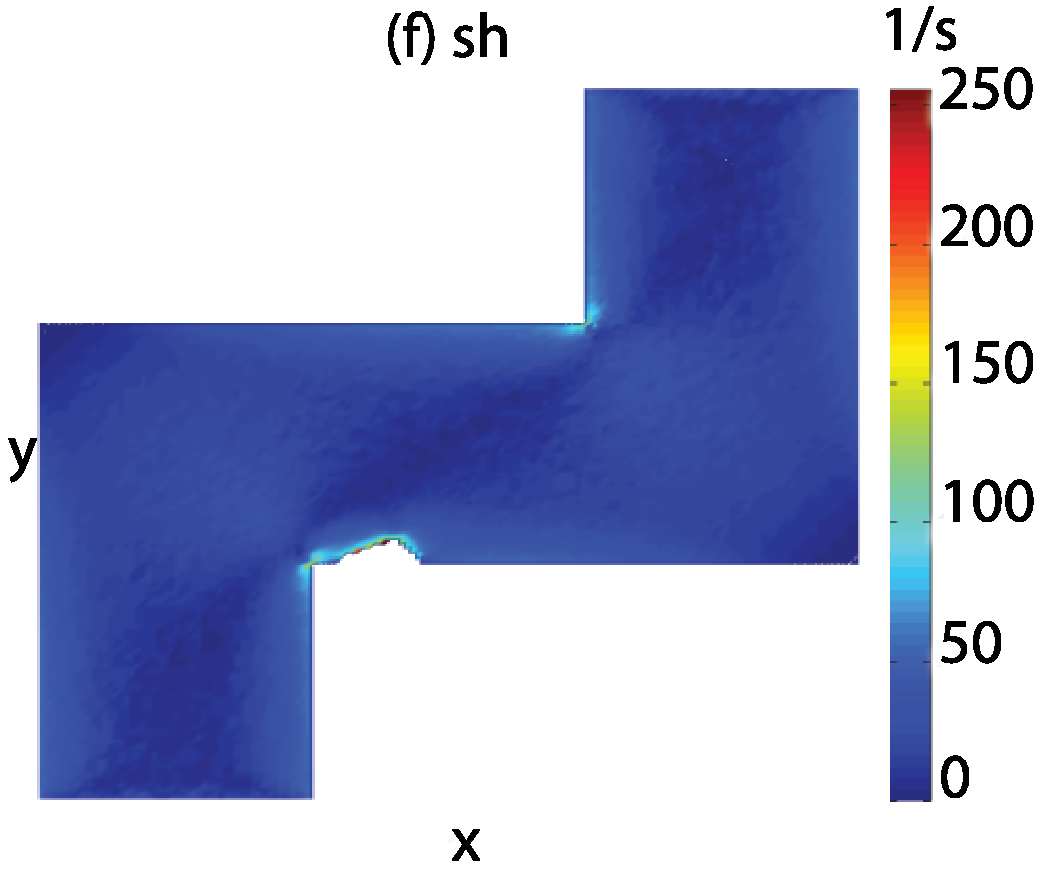}
\caption{ Initial status of a central slice $z=z_0$ of the tubes:
(a) Computational grid with biofilm seed (green), fluid (blue) and substratum (black).
(b),(c),(d) Velocity components around the initial biofilm seed. 
(e) Pressure field. (f) Shear rate.}
\label{p161:fig1}
\end{figure}

Let us denote by $\Omega_f$ the region occupied by fluid and by $\Omega_b$ 
the region occupied by biofilm. The whole computational region is divided in
a grid of tiles. Each tile may be filled with either substratum, fluid, or biomass, as 
illustrated in Fig.~\ref{p161:fig1}. Since we have in mind applications to microflows, we choose the size of each tile to be of the order of the average size of one bacterium, about $1$-$2$ $\mu \,m$.

The fluid surrounding the biofilm is governed by the incompressible 
Navier-Stokes equations:
\begin{eqnarray}
\rho {\mathbf u}_t - \mu \Delta {\mathbf u} + {\mathbf u} \cdot \nabla{\mathbf u}
+ \nabla p = 0,  & \quad {\mathbf x} \in \Omega_f, t > 0 \label{p161:ns} \\
{\rm div\,} {\mathbf u} = 0, &  \quad {\mathbf x} \in \Omega_f, t > 0 \nonumber
\end{eqnarray}
where ${\mathbf u}({\mathbf x},t)$  is the velocity and $p({\mathbf x},t)$ the 
pressure. $\rho$ and $\mu$ stand for the density and viscosity of the fluid.
The non-slip condition on the velocity holds at the biofilm/fluid interface
$\Gamma$.  A low cost prediction of the evolution of the velocity and pressure 
fields is provided by second order slight artificial compressibility schemes \cite{chorin}. Approximated velocities and pressures can be improved using 
second order implicit gauge schemes \cite{wegauge}, if necessary, at a 
higher cost. 

Flow effects are felt by a biofilm on much shorter time scales (seconds) than growth effects (hours) \cite{stonepnas}.  
Biomass attaches, detaches and moves  according to the flow fields at each location. Floating bacteria are carried by the fluid. The flow geometry selects 
preferential adhesion sites on the walls where biofilm seeds may be nucleated \cite{stonestreamer}. Biofilm nucleation may be successful or not depending on 
the surface nature and the bacterial strain. The flow also determines the strength 
of the biofilm \cite{stoneadhesion,stoodleycohesion}.  
Once a biofilm seed is formed, biomass accumulation is a balance between biomass increase due to adhesion or cellular processes, and loss of bacteria 
due to erosion \cite{stoodleynutrients}.
We describe below basic stochastic rules for adhesion, erosion and motion  processes, having in mind the model case of bacterial streamers in laminar corner microflows,  that will serve as a test later.  We focus on fast processes. Growth due to nutrient consumption is neglected here.


Two main adhesion processes are taken into account:
\begin{itemize}
\item Adhesion of floating cells to walls. In laminar regimes, nucleation
of biofilm seeds on the walls is often driven by the geometry. Corners or
narrowings may produce secondary flows that drive cells and particles to
the walls. Continuous adhesion of bacteria at preferential adhesion sites 
is taken care of by attaching $N_s$ cells at each step. They distribute on 
the seed, inside a limited region where the secondary flow is expected to 
be relevant. 
\item Once a biofilm seed sticks out from the wall, bacteria and 
particles swimming with the flow may hit it, and stick to it at a certain rate.
Additional $N_b$ biomass blocks are distributed between the tiles located at 
the biofilm/fluid interface.
\end{itemize}

$N_s$ and $N_b$ depend on the density of biomass floating in the fluid. $N_s$
is affected by the likeliness of the specific bacterial strain selected to adhere to the 
walls.


Biomass tiles ${\cal C}$ located on the surface of the biofilm detach due to shear forces exerted by the flow \cite{stoodleynutrients}. A probability for biomass detachment is proposed in \cite{hermanovic}:
\begin{eqnarray}
P_e({\cal C}) = {1 \over 1 + { \gamma \over   \tau({\cal C})}}
=  {\tau({\cal C}) \over \tau({\cal C}) +  \gamma}.    
\label{p161:erosion}
\end{eqnarray}
$\gamma$ is a measure of the biofilm cohesion. We assume it to be known  and constant.
$\tau({\cal C})$ measures the shear force felt by cell ${\cal C}$. Here,
we use the magnitude of the shear force due to the flow at the cell location $\tau_f({\cal C}) $, modified by a geometrical factor $f({\cal C})$ that accounts for the local sheltering role of neighboring cells, see \cite{pre}.
In our numerical experiments, $\tau_f({\cal C})$  is usually set equal to the shear rate at
location ${\cal C}$ multiplied by the fluid viscosity $\mu$. The shear rate is defined as the 
spatial rate of change in the fluid velocity field \cite{shearrate}.
As for the geometrical factor, it varies according to the main component of the flow, see \cite{pre}.  In practice, we check erosion in the three directions.
At each step and for each biomass tile ${\cal C}$ on the biofilm boundary, we detach
biomass with probability $P_e({\cal C})$.
Erosion due to the flow may occur as detachment of single blocks or of whole clusters of biomass with a thinning connection to the rest of the biofilm.  


Shear forces exerted by the flow on the biofilm surface detach biomass.
Normal forces on biofilm surfaces may move them. The motion of a biofilm block may be seen as the result of the collective motion of small fragments of the 
aggregate.

The probability for biomass motion in the $x$ directions is  defined as:
\begin{eqnarray}
P_x({\cal C}) = {1 \over 1 + { \gamma \over   |F_x({\cal C})| }}
=  { |F_x({\cal C})| \over |F_x({\cal C})| +  \gamma }.    \label{p161:motionx}
\end{eqnarray}
Similar expressions are used in the $y$ and $z$ directions.
$\gamma$ is again a measure of the biofilm cohesion. $F_x$ is the
force exerted by the flow in the $x$ direction (on cell walls normal
to the $x$ direction) weighted with a geometrical factor accounting
for neighbor protection similar to the one used in (\ref{p161:erosion}), \cite{pre}.
$F_y$ and $F_z$ are its counterparts in the $y$ and $z$ direction. 
The forces are calculated using the values of the fluid stress tensor $\sigma$
at the  cell location: $\sigma \cdot {\mathbf n}$ for the chosen normal
vector ${\mathbf n}$.

At each step and for each occupied tile on the biofilm boundary, the biomass moves in the $x$ direction with probability $P_x({\cal C})$ pushing its 
neighbors in that direction too. 
Motion is in the positive or negative sense depending on the sign of $F_x$. 
Similar rules are applied in the $y$ and $z$ directions.    

\section{Numerical results}
\label{p161_num}

We will fix as a model case of study the growth of streamers in corner microflows, 
that is well documented experimentally \cite{stonestreamer}.
The computational region is described in Fig.~\ref{p161:fig1}(a). A pressure 
driven flow circulates through the ducts with maximum velocities of about $1$
$mm/s$. The structure of the flow is represented in Figs.~\ref{p161:fig1}(b)-(f).
The density of the liquid is $10^3$ ${Kg\over m^3}$ and its viscosity 
$\mu=10^{-3}$ $Pa \cdot s$. The bacterial size, and the tile size thereof, is 
taken to  be $2 \, \mu \, m$. The dimensions of the central straight  fragment 
are $N\times M  \times L $ $\mu \, m$. Streamers grow mostly in the 
$N/3 \times M \times L $ $\mu \, m$ region between corners.
In real experiments, usual values for $N,$ $M$ and $L$ are $600$, $200$ 
and $100$. In the numerical tests selected here, we have divided those sizes 
by $2$ to reduce the computational cost. 

An initial biofilm seed is placed on the left corner at the bottom, see Figure 
\ref{p161:fig1}(a). According to \cite{stonestreamer}, the presence of secondary
vortices in that area  favors adhesion of particles to the wall, becoming
a preferential adhesion site. Biomass will be attached to that seed,
eroded and moved according to stochastic rules described above.

\begin{figure}
\centering
\hskip 1cm (a) \hskip 5cm (b)\\
\includegraphics[width=4cm]{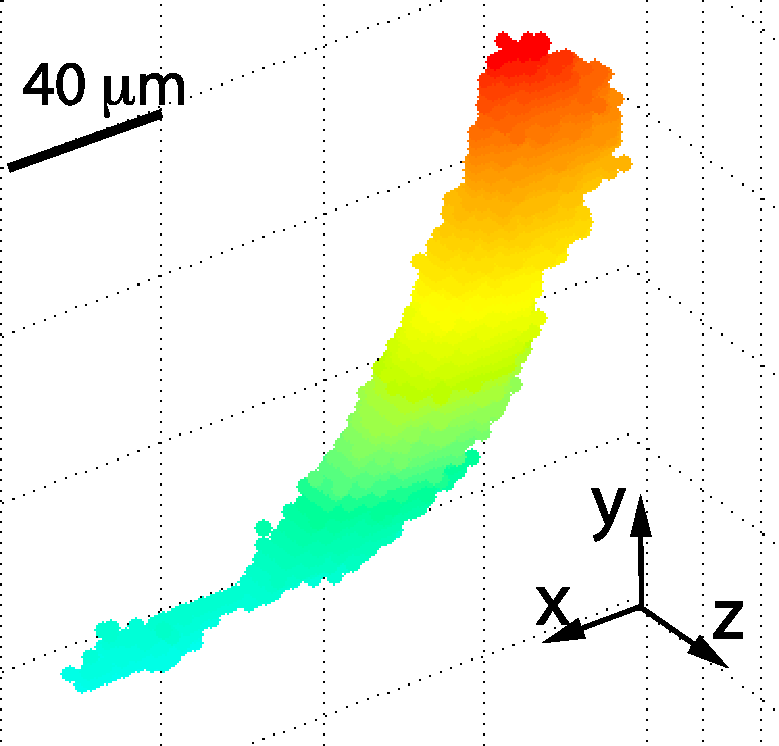} \hskip 1cm
\includegraphics[width=4cm]{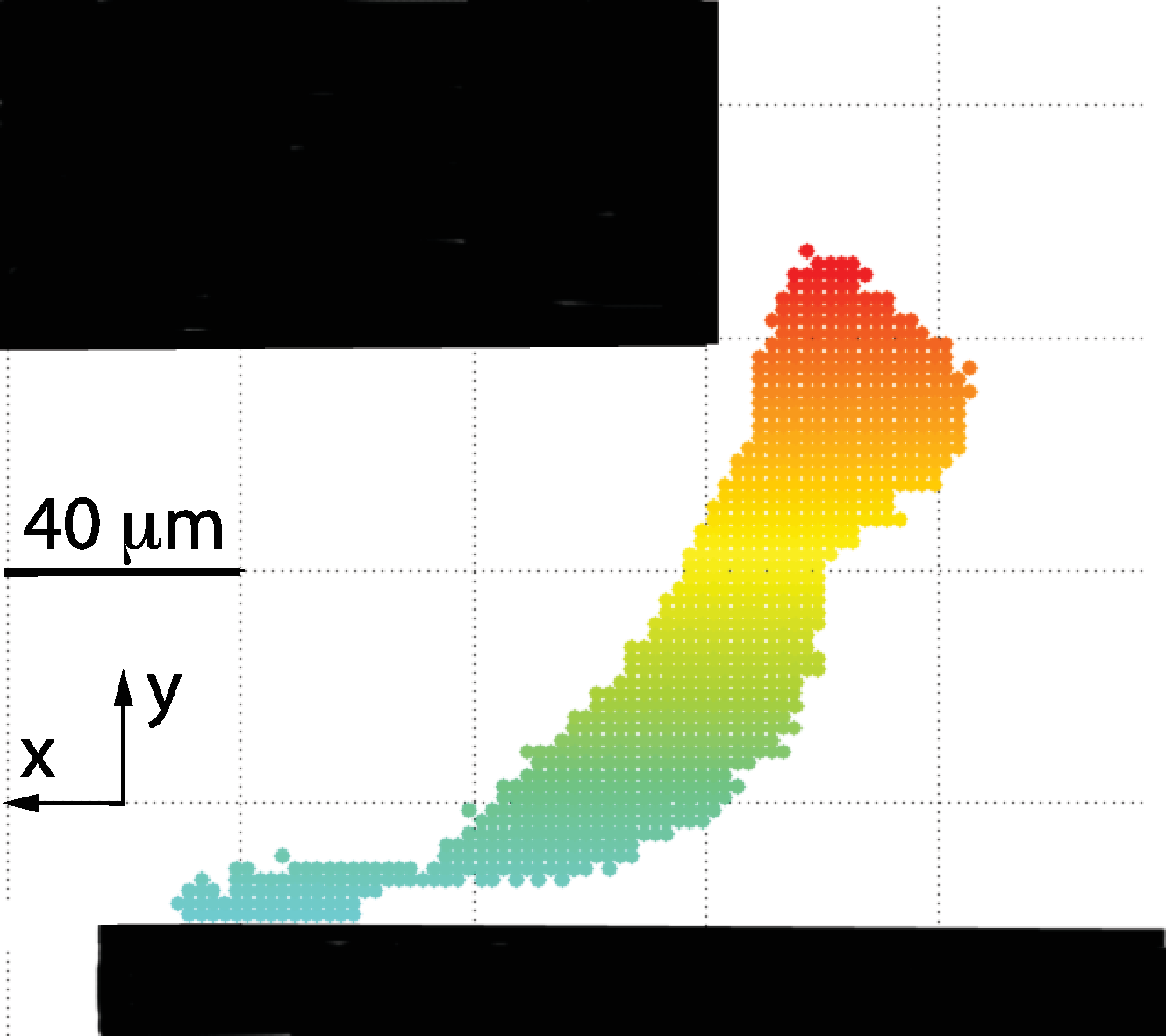}
\caption{Streamer grown for $\gamma=15 \, Pa$ at step $12600$ of the adhesion-
erosion-motion process. $N_s=1$ around the initial seed and $N_b=4$ along 
the biofilm body. The biofilm is merging with another seed growing at the 
opposite corner, which has been ignored in the plot: (a) front view, (b)
side view. }
\label{p161:fig2}
\end{figure}

Numerical tests of biofilm growth are performed using this  
geometry, see Fig.~\ref{p161:fig2}. $\gamma$ is a measure of the biofilm 
cohesion  estimated from the biofilm Young modulus. 
Reference \cite{stonestreamer} gives values in the range $70-140$ $Pa$. 
To reduce the computational cost, we adjust it so that 
our biofilms involve a small number of tiles.  Images in Reference
\cite{stonestreamer} yield estimates for the adhesion time $\tau$ 
of $1$ block of biomass per second. Each step of the 
adhesion-erosion-motion process occurs in a time scale $\tau$.  

Provided enough biomass attaches to the seed (to avoid streamer detachment) and to the biofilm body (to resist increasing erosion while crossing the 
current), the aggregate grows into the current, elongates with it, bends when
it reaches the curve, approaches the opposite corner, and eventually merges with 
the additional biofilm seed that should be growing there.  The observed effective growth rate is the balance between the biomass that attaches and detaches at each step, and varies during the spread process. It is usually larger before the thread tries to cross the main stream and decreases as it tries to reach the opposite corner while changing its shape. 

The aggregate grows into the region of minimum shear rate, that joins the two
corners. Once formed, pressure variations move the filament downstream, 
curving it in a similar way to the experimentally observed threads, and
leaving a thin joint with the seed. It reaches the opposite corner from
behind, as observed in experimental photographs. 

\begin{figure}
\centering
\hskip 0cm (a) \hskip 3cm (b) \hskip 3cm (c) \\
\includegraphics[width=3.8cm]{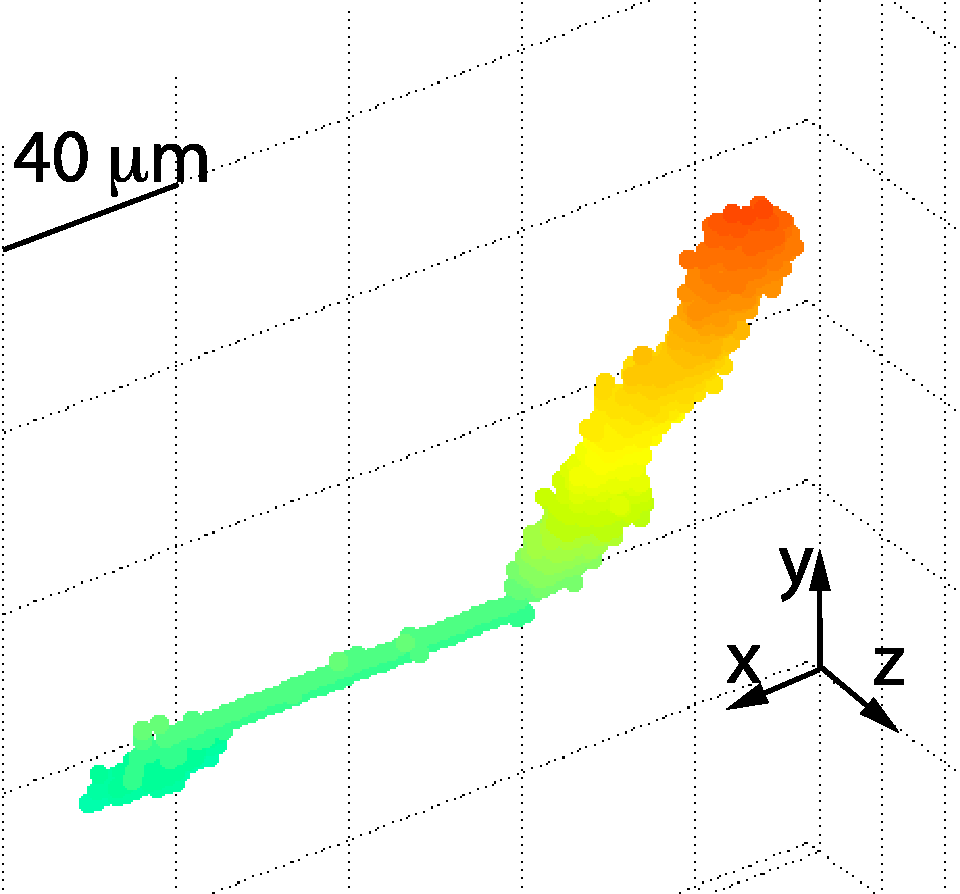} \hskip 2mm
\includegraphics[width=3.4cm]{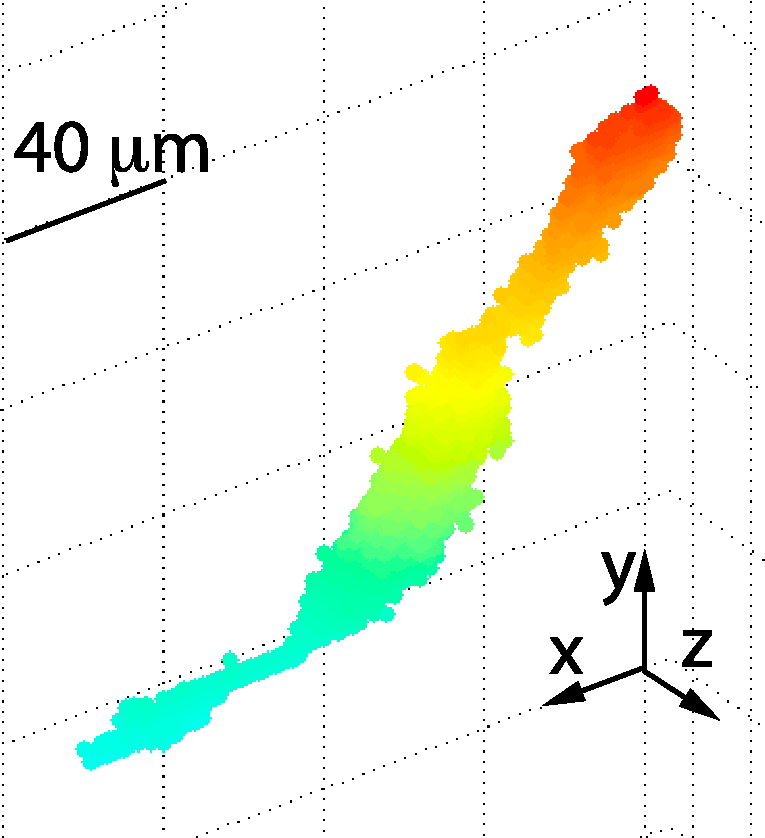} \hskip 2mm
\includegraphics[width=3.5cm]{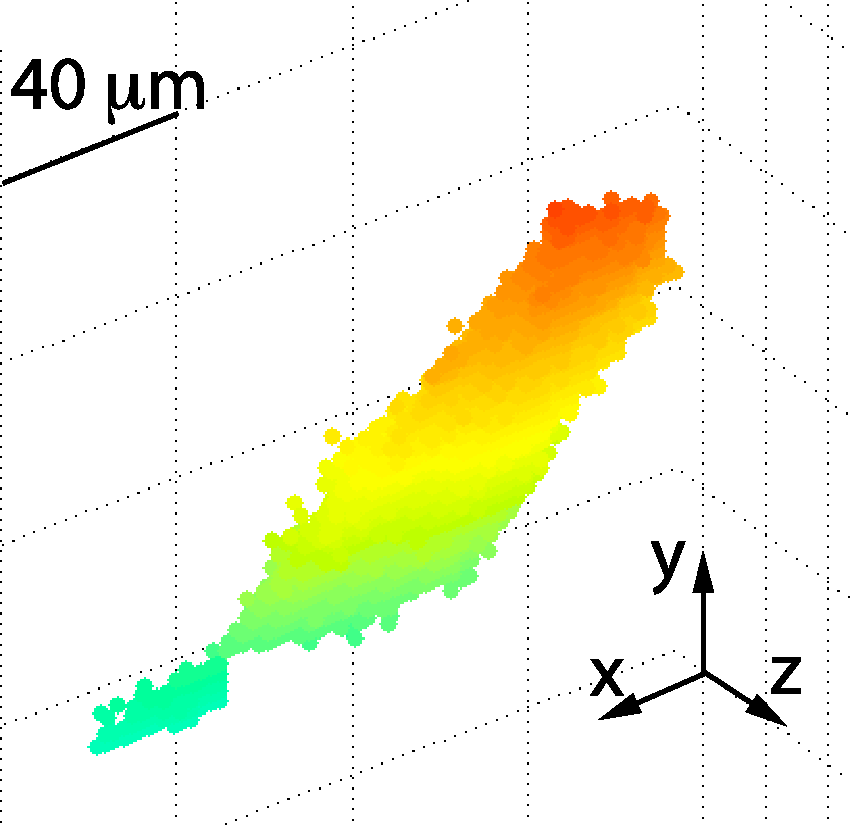}
\caption{Reducing the number of attached biomass blocks, streamers
detach without reaching the opposite corner. 
(a) Decreasing $N_b$ to $2$, the streamer elongates, bends, detaches
and regrows. The image corresponds to step $42600$, just before the
fourth detachment, with $1373$ blocks.
(b) Decreasing $N_b$ to $3$, the streamer becomes too thin and the top
part encounters resistance to join the corner. It finally breaks off at step
$15600$, with $2151$ blocks. 
(c) Decreasing $N_s$  to $0.5$ (one block attached each two steps), the 
connection of the streamer to the seed breaks off after step $9700$ with
$4792$ blocks.
Other parameter values as in Figure \ref{p161:fig2}. Distance between
grid lines is always $40 \mu m.$
}
\label{p161:fig3}
\end{figure}

\begin{figure}
\centering
\hskip 0cm (a) \hskip 3cm (b) \hskip 3cm (c) \\
\includegraphics[width=3.5cm]{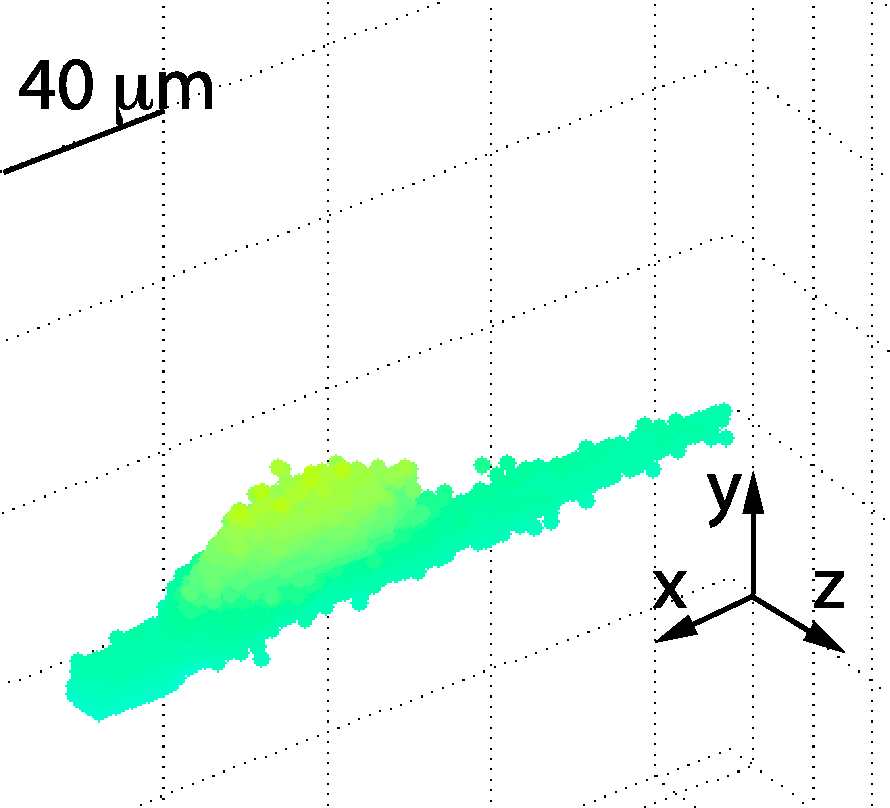}
\includegraphics[width=3.5cm]{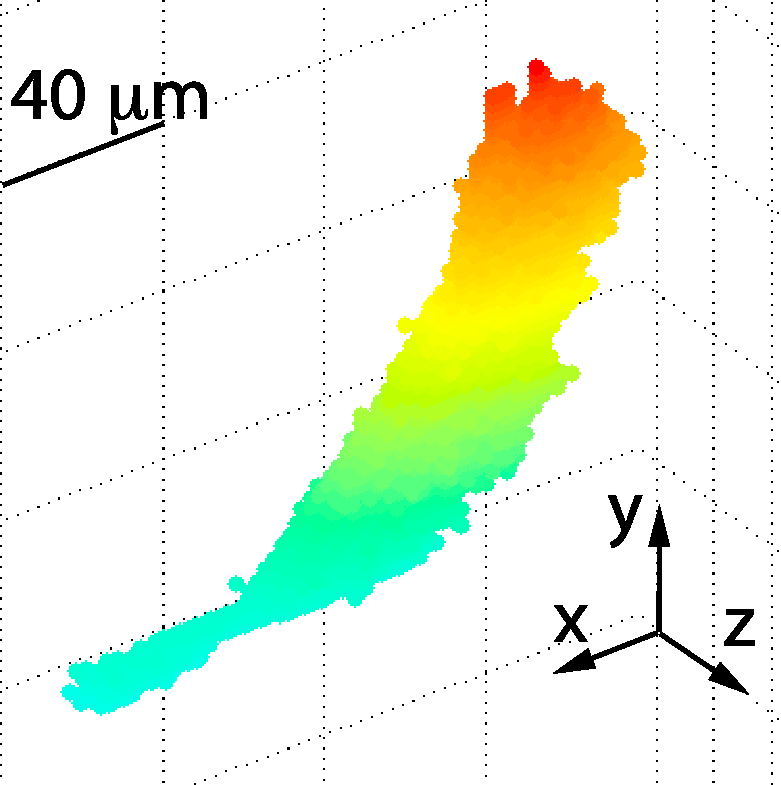}
\includegraphics[width=3.5cm]{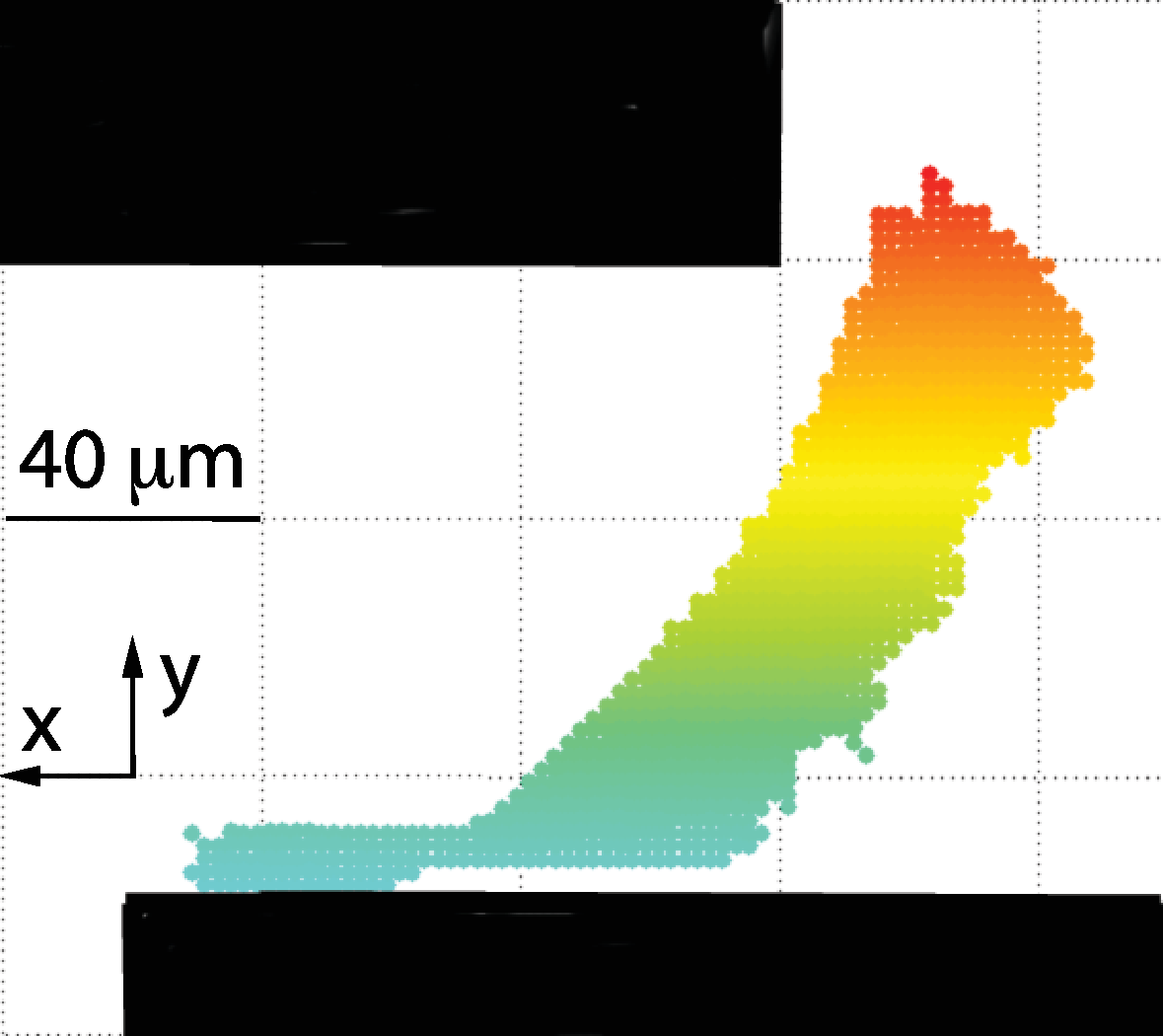}
\caption{
(a) Increasing $N_s$ to $2$ the streamer  remains  parallel to the substratum 
until it reaches the  wall at step $3200$ with $3242$ biomass blocks, for 
$\gamma=15 \, Pa$ and $N_b=3$. 
Increasing $\gamma$ to $20\, Pa$,  the thread widens and crosses the current. 
(b) and (c) show the front and lateral views at step $15000$, with $4702$ 
blocks.
}
\label{p161:fig4}
\end{figure}

The number of biomass blocks to be attached depends on the selected biofilm cohesion. Too large values of $N_b$ produce expanding balls. Too small adhesion rates to the biofilm $N_b$ produce an elongated thread close to the wall, that eventually feels the corner flow and starts to gain biomass on the top, but may not receive enough biomass to resist the increased erosion and  detaches, see Figures \ref{p161:fig3} (a) and (b). For small values of $N_s$ the connection between the streamer and the seed breaks off, see Figure \ref{p161:fig3} (c). Too large adhesion rates to the seed $N_s$ favor expansion parallel to the bottom substratum. If $N_b$ is not large enough for the selected cohesion, the biofilm reaches the rightmost wall as shown in Figure \ref{p161:fig4} (a). 
Increasing $N_b$, the biofilm may cross to the opposite corner sustained by a wider basis.  If the initial adhesion rates are large enough for the considered cohesion, a sort of fan expands into the main stream.  The fan becomes narrower as we reduce the adhesion rates. 

Depending on the ratio  $N_b/N_s$ for the selected $\gamma$, we see narrower or wider streamers. If we increase the cohesion parameter $\gamma$, we must reduce the computational adhesion rates $N_b$ and $N_s$ to see similar behaviors. The failed streamer in Figure \ref{p161:fig4} (a)
reaches successfully the opposite corner sustained by a wider basis when
we slightly increase $\gamma$ in Figure \ref{p161:fig4} (b), (c).
If the biofilm cohesion is too small, the biofilm seed is eroded and eventually 
washed out.  No thread is formed. 

These tests provide insight on the way these structures are formed. Threads experimentally observed  \cite{stonestreamer}, however, look more like thin jets
and may require a different description. Streamers joining opposite corners appear to be attractor shapes that may be formed under different dynamics.

\vskip 2mm
{\bf Acknowledgement.}
D. R. Espeso and A. Carpio were supported by the Autonomous Region of 
Madrid and the spanish MICINN through grants S2009/ ENE-1597, 
and FIS2011-28838-C02-02. 
B. Einarsson was supported by a grant of the NILS program and project FIS2008-04921-C02-01. 

%
%



\end{document}